\documentclass[sigconf]{acmart}
 \pdfoutput=1

\renewcommand\footnotetextcopyrightpermission[1]{} 
\makeatletter
\renewcommand\@formatdoi[1]{\ignorespaces}
\makeatother
\usepackage{comment}
\usepackage{xcolor}
\usepackage{listings}
\usepackage{graphicx}
\usepackage{natbib}
\usepackage{subfiles}
\definecolor{codegreen}{rgb}{0,0.6,0}
\definecolor{codegray}{rgb}{0.5,0.5,0.5}
\definecolor{codepurple}{rgb}{0.58,0,0.82}
\definecolor{backcolour}{rgb}{0.95,0.95,0.92}

\lstdefinestyle{mystyle}{
  backgroundcolor=\color{backcolour},   commentstyle=\color{codegreen},
  keywordstyle=\color{magenta},
  numberstyle=\tiny\color{codegray},
  stringstyle=\color{codepurple},
  basicstyle=\ttfamily\footnotesize,
  breakatwhitespace=false,         
  breaklines=true,                 
  captionpos=b,                    
  keepspaces=true,                 
  numbers=left,                    
  numbersep=5pt,                  
  showspaces=false,                
  showstringspaces=false,
  showtabs=false,                  
  tabsize=2
}

\begin{document}

\title{Verifying Software Vulnerabilities in IoT Cryptographic Protocols}

\author{Fatimah Aljaafari}
\orcid{000-0001-7683-8182}
\affiliation{%
  \institution{The University of Manchester}
  \city{Manchester}
  \state{UK}
}

\author{Lucas C. Cordeiro}
\affiliation{%
  \institution{The University of Manchester}
  \city{Manchester}
  \country{UK}}

\author{Mustafa A. Mustafa}
\affiliation{%
  \institution{The University of Manchester}
  \city{Manchester}
  \country{UK}
}


\renewcommand{\shortauthors}{Aljaafari, et al.}

\begin{abstract}
Internet of Things (IoT) is a system that consists of a large number of smart devices connected through a network. The number of these devices is increasing rapidly, which creates a massive and complex network with a vast amount of data communicated over that network. One way to protect this data in transit, i.e., to achieve \textit{data confidentiality}, is to use lightweight encryption algorithms for IoT protocols. However, the design and implementation of such protocols is an error-prone task; flaws in the implementation can lead to devastating security vulnerabilities. These vulnerabilities can be exploited by an attacker and affect users' privacy. There exist various techniques to verify software and detect vulnerabilities. Bounded Model Checking (BMC) and Fuzzing are useful techniques to check the correctness of a software system concerning its specifications. Here we describe a framework called Encryption-BMC and Fuzzing (\textit{EBF}) using combined BMC and fuzzing techniques. We evaluate the application of \textit{EBF} verification framework on a case study, i.e., the S-MQTT protocol, to check security vulnerabilities in cryptographic protocols for IoT.
\end{abstract}


\maketitle
\setlength{\parskip}{.05cm plus4mm minus3mm}
\section{Introduction}

Internet of Things (IoT) is a system that consists of a large number of smart devices and objects, such as RFID tags, sensors, actuators, and smartphones, which communicate with each other with minimum human interventions~\cite{2}.  These smart devices can connect through Wifi, Bluetooth, and RFID~\cite{2}. Additionally, these devices have embedded sensors or RFID tags to send and receive data. IoT covers different areas and applications, such as smart homes, cities, and smart health care~\cite{7}. According to Singh et al.~\cite{2}, the number of objects in IoT, excluding computers, tablets, and cellphones, will create US \$$300$ billion revenue until $2020$. Singh et al. also pointed out that by $2020$, the number of cellphones and tablets will be around $7.3$ billion devices~\cite{2}. These objects will create a massive and complex network with a considerable amount of data communicated over this network~\cite{11,12}. The existence of such a vast network of connected devices will pose new \textit{security}, \textit{privacy}, and \textit{trust} issues, which put users at high risk~\cite{2}.

One of the ways to protect data in transit, thus achieve \textit{data confidentiality}, is by designing bespoke lightweight encryption algorithms for IoT devices~\cite{14}. Due to limitations in IoT devices such as power consumption, low memory life, low processing speed~\cite{29,30}, lightweight encryption algorithms, such as PRESENT \cite{34}, Hummingbird \cite{35}, and WG-8~\cite{36}, have different properties than traditional encryption algorithms such as AES~\cite{39} and RSA~\cite{37}. Thus, different types of lightweight encryption algorithms exist, which are suitable for IoT devices. These algorithms have a small key/block size and many rounds~\cite{15}. However, implementing encryption for protocols is an error-prone task~\cite{25}; flaws in the implementation can lead to devastating security vulnerabilities~\cite{25}. According to the Common Vulnerability and Exposure (CVE)~\cite{68}, the number of vulnerabilities reported in $2018$ is $16555$, more than double what was reported in $2016$ (i.e., $6447$). An attacker can exploit these security vulnerabilities and affect users' privacy. 

\vspace{1ex}

There exist various techniques for finding such vulnerabilities~\cite{16,50}. For example, fuzzing~\cite{50} is an automated software testing technique that involves providing invalid or unexpected values as inputs to a program. Another popular technique is bounded model checking (BMC)~\cite{Biere09}, which can detect bugs by using symbolic values. The basic idea of BMC is to check the negation of a given property at a given depth. However, both fuzzing and BMC have not yet been exploited to detect vulnerabilities in implementations of lightweight cryptographic protocols for IoT devices. In particular, performing fuzzing and BMC alone is inadequate in achieving high-path coverage efficiently. BMC struggles in reaching deep parts of the code because of the state-space explosion issue and its dependency on Boolean Satisfiability (SAT)~\cite{25} or Satisfiability Modulo Theories (SMT) solvers~\cite{69}. Fuzzing also has its drawbacks as it is unable to examine all code paths~\cite{27}. 

\vspace{1ex}

We make two significant contributions towards the verification of cryptographic protocols in IoT devices. First, we combine fuzzing and BMC in an unprecedented manner to increase code coverage and to detect a maximum number of security vulnerabilities of cryptographic protocols in IoT. Precisely, we execute the program under test using a BMC engine for a particular amount of time. After the first round of BMC, we aim to cover all the functions through various inputs. However, due to limitations in the underlying solvers, BMC can only reach shallow coverage (i.e., it never reaches branches that lie deep in the code)~\cite{28}. To solve the issue of the shallow coverage of BMC, as the next step of our technique, we use the inputs or packets in case of protocols generated by BMC as seed inputs for the fuzzer.
In contrast to Ognawala et al.~\cite{28} who combine symbolic execution and apply fuzzing to general-purpose software, our proposed method starts with BMC and then use fuzzing by considering intricate properties in IoT protocols. Second, we perform a preliminary evaluation of the application of the Encryption-BMC and Fuzzing (\textit{ESF}) verification framework to detect security vulnerabilities in implementations of cryptographic protocols for IoT. Although our \textit{EBF} prototype is not entirely implemented, its application to go beyond design correctness and also to address behavior correctness represents an important research direction in ensuring secure IoT protocols.

\section{Preliminaries}
\subsection{Lightweight cryptographic protocols for IoT devices}

Lightweight encryption is suggested as a concept to handle security in resource constraint devices such as IoT devices~\cite{47}. This lightweight encryption is based on the traditional principles of cryptography, but its purpose is to design algorithms, which have stricter requirements than other algorithms in terms of performance. They must be ultra-efficient as they will be run on devices with low computational capabilities~\cite{47}. Thus, they can be used in the IoT protocols to secure communication between two endpoints. Encryption schemes can be divided into two main categories: 
(1) \textit{Symmetric key encryption}, which uses the same key for both encryption and decryption~\cite{15} such as AES~\cite{39}, DES and 3DES~\cite{38}; and (2) \textit{Asymmetric key encryption}, which uses different keys, a public key for encryption and private key for decryption~\cite{15}, such as RSA~\cite{56} and Elliptic Curve Cryptography (ECC)~\cite{47}.
We evaluate our \textit{EBF} prototype on the Secure Message Queue Telemetry Transport (S-MQTT) protocol~\cite{22}. S-MQTT is a messaging transport protocol for IoT devices, which uses Transport Layer Security (TLS)~\cite{70} that support different types of cipher suite such as AES.


\subsection{Software verification for lightweight cryptographic protocols}

Here we focus on verifying lightweight cryptographic protocols for IoT devices using fuzzing and BMC to detect security vulnerabilities. Given the current knowledge in software verification, our proposed verification algorithm is the first to address the implementation aspects of cryptographic protocols for IoT devices, such as memory leak and buffer overflow. In implementations of lightweight cryptographic protocols, a memory leak may occur if the developer unintentionally fails to free an assigned block of memory when that chunk of memory is no longer needed~\cite{48}. Another typical implementation error is buffer overflow, which happens when a program attempts to write more data and exceeds the length of the allocated memory. In any particular case, the attacker then can exploit these vulnerabilities and affect the system. For example, Cisco found a vulnerability in the cryptographic driver for Adaptive Security Appliance Software (ASA) and Firepower Threat Defense (FTD) Software~\cite{72}. This vulnerability occurred because of the insufficient input validation of a Secure Sockets Layer (SSL) or TLS packet header. It allows the attacker to reboot the device suddenly, which will result in a denial of service (DoS) condition~\cite{Cisco2018}. 

\subsubsection{Fuzzing}

Fuzzing is an automated security vulnerabilities detector technique, which works by feeding protocol implementations with random or semi-random inputs~\cite{50}. The specification of the software does not foresee these inputs. Then, the system behavior is checked for abnormalities, such as crashes or failures.~\cite{27}. Fuzzing often suffers from a low code coverage dilemma~\cite{51}. Fuzzing can detect implementation level flaws, but it cannot find design issues. For example, fuzzing can detect buffer overflow, but it cannot detect the use of weak encryption algorithm~\cite{52}. In terms of cryptographic protocols, we can fuzz a running protocol that uses encryption with invalid packets and check the connection behavior of the client and server. In our proposed verification method, we provide to our fuzzer the necessary cryptographic keys/algorithms in a log file.

\subsubsection{Bounded Model Checking (BMC)}

A principal aim of BMC in the context of software verification is to examine bounded program runs and thus generate verification conditions (VCs) that reflect the exact path in which a statement is executed, the context in which a given function is called, and the bit-accurate representation of expressions~\cite{Biere09,CordeiroFM12}. A VC is a logical formula (constructed from a bounded program and desired correctness properties), whose validity implies that a program's behavior agrees with its specification. Correctness properties, in programs, can be specified by users via \textit{assert} statements or automatically generated from a specification language. If all of a bounded program's VCs are valid, then it complies with its specification, up to a given bound. For cryptographic protocols, we deal with symbolic packet instead of regular input. Our BMC engine will initially explore the cryptographic functions, which will help the fuzzer to examine the paths deeply.  

\section{Detecting Vulnerabilities in Implementations of Lightweight Cryptographic Protocols}
\label{section:ESF}

A large number of recent attacks targeting improper implementations of cryptographic protocols, such as man-in-the-middle and side-channel attacks~\cite{27}, has motivated us to provide security analysis of cryptographic protocols such as S-MQTT~\cite{22}. Here we develop a novel verification algorithm to detect vulnerabilities such as buffer overflow and memory leak using fuzzing and BMC techniques. The novelty of our approach does not rely only on a scalable and precise verification method to check for implementation errors but also in the application domain due to the importance of securing IoT and its respective lack of research on finding security vulnerabilities in cryptographic protocol implementations. Therefore, the main objective of our proposed method is to overcome the limitations of fuzzing and BMC in isolation, such as low-code coverage and state-space explosion resp., by combining them to achieve precision and scalability in an unprecedented manner. We aim to apply our method to ensure the safety and security of IoT software and also evaluate the effectiveness of this method by determining the code coverage and the number of found vulnerabilities. Lastly, we aim to analyze and compare the results with other verification tools such as Csur~\cite{53} and CDF~\cite{56}, which are designed to verify traditional cryptographic protocols. As a result, our ultimate goal is to detect implementation and behavior errors in cryptographic protocol implementations by combining fuzzing and BMC, thus improving safety and security in IoT devices.

\subsection{Encryption-BMC and Fuzzing (\textit{EBF}) Method}

Here we describe our \textit{EBF} verification method, which combines fuzzing with BMC in an unprecedented manner. Our \textit{EBF} method can be described as follows:  

\begin{enumerate}
    \item We symbolically execute the initial state of the cryptographic protocol implementation (e.g., the S-MQTT implementation). Since we are dealing with protocols, we will use packets instead of regular inputs to our BMC engine to produce test cases that can exercise different program paths. 
    \item We expect to cover most program paths using symbolic packets after the first round of the BMC engine. To create symbolic packets, we will replace some bytes of concrete packets by symbolic values.
    \item We use the inputs generated by BMC as seed inputs for the fuzzing tool in the form of a test case. These test cases contain symbolic values, but we do not expect them to reveal any vulnerabilities. These test cases will be regenerated with concrete values extracted from the BMC engine, to feed them to the fuzzer engine for the second round.
    \item We start a protocol session between two endpoints to share encrypted messages (which is the important part we want to test in the fuzzing stage) and pass the test cases to the running protocol to check for irregularity, e.g., denial-of-service. The server will also generate inputs and pass them to the fuzzer to ensure the exploration of all the paths. Then the fuzzer will mutate the inputs and send them to the client.
    \item Our fuzzer reads the required keys and cryptographic algorithms to mutate encrypted messages properly from a log file to explore different unique paths, which were initially difficult to reach. This particular log file will be automatically exported from the server implementation; it will contain the key and the encryption algorithm used.
    \item If the fuzzer explored all the paths that BMC had difficulty to cover and used all the inputs generated from BMC, then it can collect and save the results within the log file.
\end{enumerate}

\begin{figure}
	\centering
	\vspace{3ex}%
	\includegraphics[width=\columnwidth]{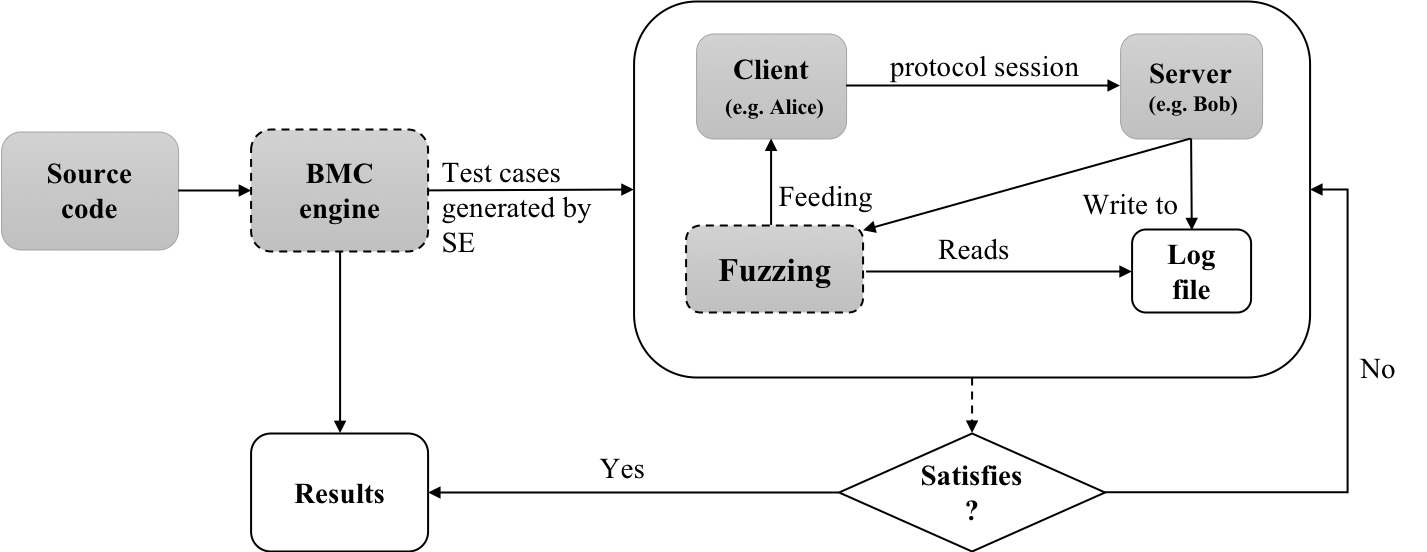}
	\caption{\textit{EBF}: overall methodology.}
	\label{fig:overall-methodology}
\end{figure}

Figure~\ref{fig:overall-methodology} illustrates our \textit{EBF} verification method.
We will build our \textit{EBF} verification framework on top of two tools to implement our methodology; these tools have been chosen based on the comparison by Beyer et al.~\cite{73} between testing and model checking. In the BMC phase, we will use the ESBMC tool for the initial state exploration to search for various implementation errors, including memory leak and buffer overflow. We then will use the AFL tool for the fuzzing phase to complete the verification by testing running sessions for the cryptographic protocol. In order to provide the fuzzer with the necessary keys and encryption algorithm, we will develop a file, which the fuzzer can read such information from and also implement a code to analyze the verification results. 

\subsubsection{Illustrative Example}

As an illustrative example, we describe our part of the EBF verification framework using the S-MQTT protocol. In the first step of our method, we use our BMC engine to check for errors in the S-MQTT implementation for a short amount of time. Then we check whether the engine finds some vulnerabilities by collecting the log results. After that, we take the test cases generated by the BMC phase and feed them to the fuzzer as the second step. Fuzzer will use the test cases as input. Then we execute the protocol and feed the client with the invalid inputs and check its behavior. The client communicates with the server using encrypted messages; if the server decrypts messages correctly, then it passes the encryption key to the log file. Also, the server will generate invalid inputs and pass them to the fuzzer to ensure the exploration of all the paths, particularly the paths BMC could not reach. By combining both techniques, we expect to cover most paths in the code.

\subsubsection{Preliminary Evaluation}

\begin{figure}
	\centering
	\vspace{3ex}%
	\includegraphics[width=\columnwidth]{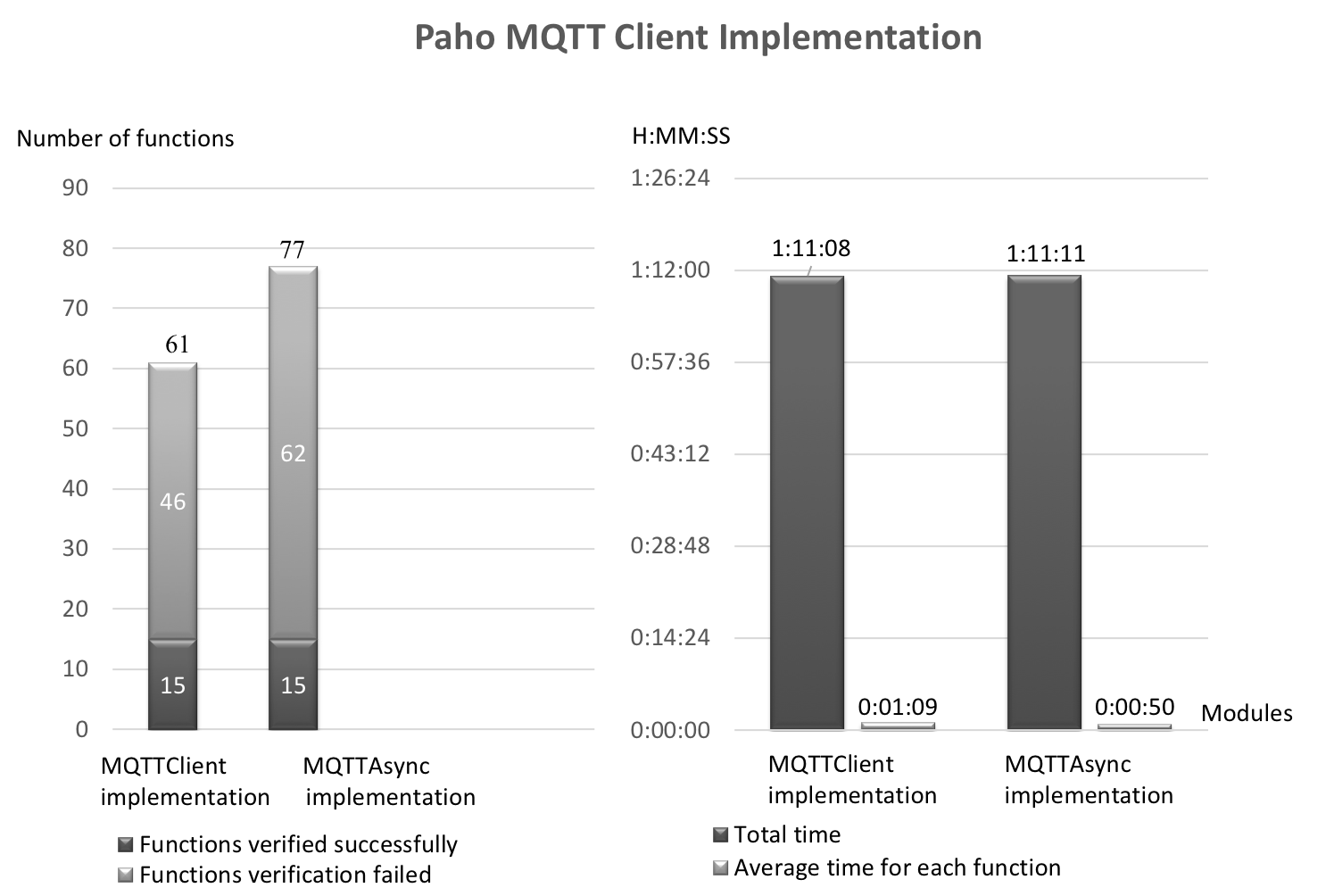}
	\caption{Numbers of functions tested and verification time.}
	\label{fig:Function_comparision}
\end{figure}

Here we only will evaluate two parts of our EBF methodology. First, we will evaluate the BMC engine as illustrated in Figure~\ref{fig:overall-methodology} by evaluating the ESBMC tool~\cite{78}. Second, we will evaluate fuzzing by evaluating AFL tool ~\cite{79}. We used Paho MQTT client implementation V1.3.1 (Synchronous and Asynchronous modes)~\cite{76} and Mosquitto MQTT server V1.6.7~\cite{77}, which support TLS V1.3. The main objective of this initial evaluation is to check whether the ESBMC and AFL tools are promising tools to be further developed in terms of verifying cryptographic protocols.
 The first implementation, which is the synchronous mode, consists of 61 functions, and the second one is an asynchronous mode, which consists of 77 functions. In Figure~\ref{fig:Function_comparision}, we can see the number of functions, which were verified successfully, and which functions ESBMC detected a bug in them. Most bugs found were array bounds and invalid pointer; as future work, we will validate these bugs based on witness checking, as part of our methodology, and also confirm these bugs with the developers.  However, regarding the time, the entire implementation required around one hour to be tested by ESBMC. Thus, the average time of each function in both modes was around one second, as shown in figure~\ref{fig:Function_comparision}. 
 
\begin{figure}
	\centering
	\vspace{3ex}%
	\includegraphics[width=\columnwidth]{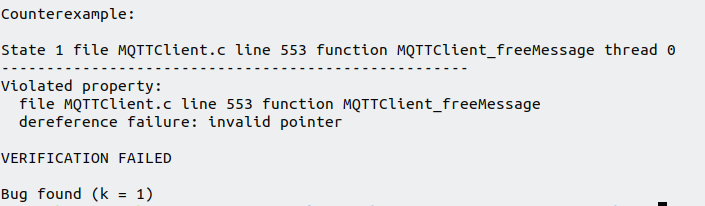}
	\caption{Counterexample for function MQTTClient freeMessage.}
	\label{fig:counter}
\end{figure}

\begin{figure}
	\centering
	\includegraphics[width=\columnwidth]{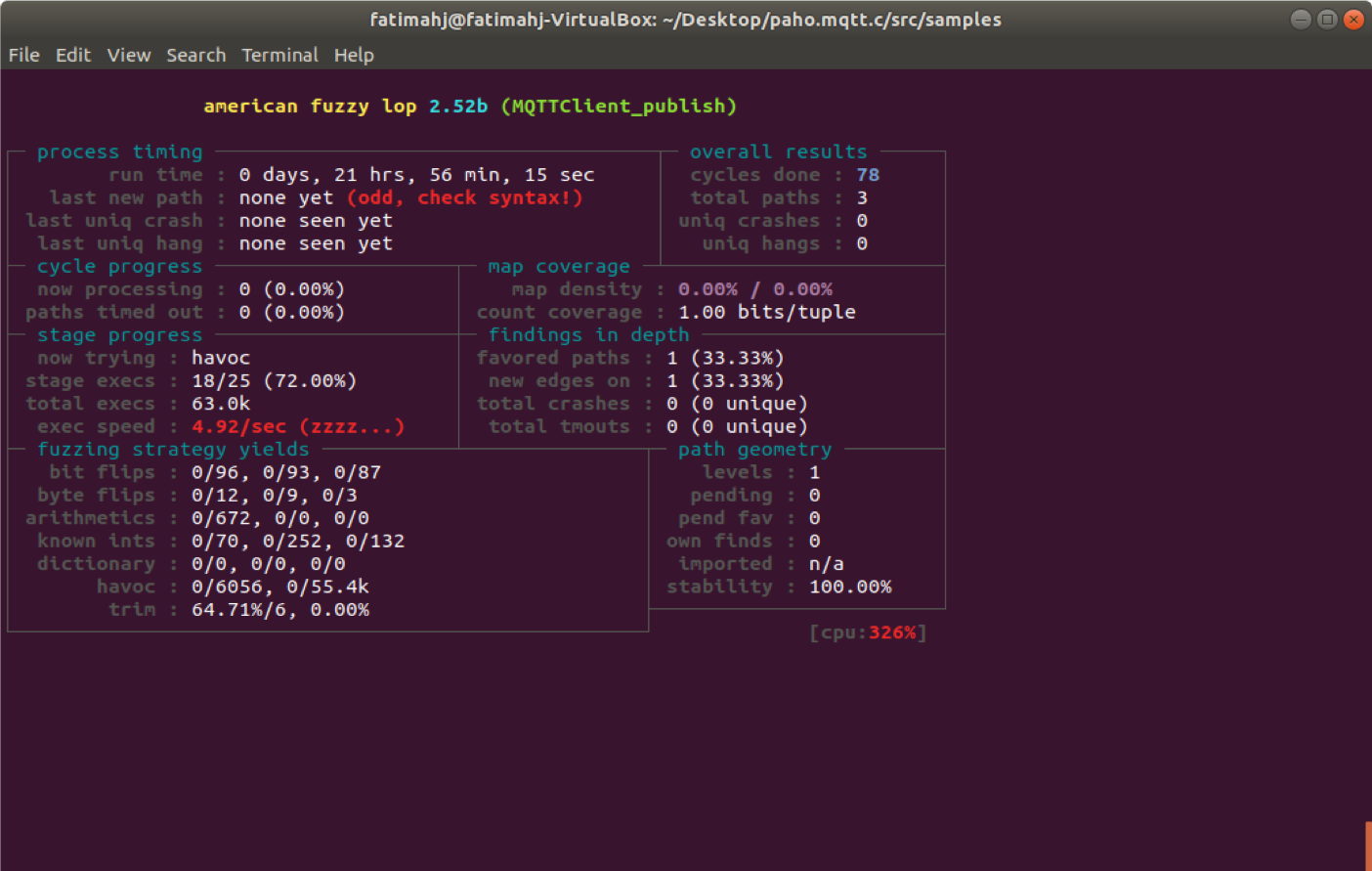}
	\caption{Fuzzing MQTT protocol session between publisher (client) and broker (server) using AFL.}
	\label{fig:AFL}
\end{figure}

 As example, we considered a simple code fragment of function $MQTTClient$\_$freeMessage$ in MQTT synchronous mode implementation, this function frees the memory allocated to an MQTT message, including the additional memory allocated to the message payload. However, the developer of this function does not check whether the pointer passed as argument to \textit{$MQTTClient$\_$message$} points to an a live object. Figure~\ref{fig:counter} illustrates a fragment of the counterexample produced by ESBMC for the function in Listing 1. One way to fix this pointer dereference failure is to ensure that \textit{message} is different than NULL. 
 As the next step in our EBF method, we initially evaluate AFL. We manually generated three test cases as a start and fed them to the tool. Unfortunately, AFL required a long time to find crashes, e.g., almost one day running, and still, did not find any crashes (cf. Figure~\ref{fig:AFL}). AFL needs further evaluation and development in terms of cryptographic protocols in the next stage of our research.  
 


\lstset{style=mystyle}

\begin{lstlisting}[language=c, caption=Fragment of code of MQTTClient]
void MQTTClient_freeMessage(MQTTClient_message** message)
{
	FUNC_ENTRY;
	MQTTProperties_free(&(*message)->properties);
	free((*message)->payload);
	free(*message);
	*message = NULL;
	FUNC_EXIT;
}

\end{lstlisting}

\section{Limitations, Related Work and Future Directions}


Many IoT applications depend on complicated cryptographic protocols for communicating over an insecure Internet. The C programming language is mostly used in writing cryptographic software~\cite{26}. Both the design and implementation of cryptographic primitives such as encryption and hash functions are error-prone. However, flaws in the implementation can lead to devastating security vulnerabilities, which may be exploited by an attacker~\cite{25}. Recent years have seen a real development in software verification of cryptographic protocols, as witnessed by the development of different tools either in BMC or fuzzing techniques~\cite{26}. However, there exists still a need for further development of these tools. Although the obligation to verify cryptographic protocols is now well identified, a few recent studies try to suggest solutions. 

One of the attempts to verify cryptographic primitive using symbolic execution is Csur~\cite{53}, which analyzes C code for security properties using a set of horn clauses, which are then solved using a theorem prover. One of the limitations of this method is that the results are only valid in a symbolic model of cryptography proposed by Dolev and Yao~\cite{64}. Also, Vanhoef and Piessens~\cite{25} modified the KLEE tool to efficiently handle cryptographic protocol by simulating their behavior under the Dolev-Yao model. Aizatulin et al.~\cite{26,63} combined symbolic execution with proof techniques by using it to obtain a ProVerif model from a cryptographic protocol implementation in C. However, the limitation of their technique is that it only considers a single execution path. Another tool used to verify cryptographic primitives using fuzzing is CDF~\cite{56}, which can be used to achieve security verification, and in particular, to find logic bugs with standard specifications. It uses differential fuzzing technique, to find inconsistencies between two implementations of the same primitive, e.g., between two implementations of RSA cipher~\cite{56}. One of the weaknesses of CDF is that it provides limited detection of timing leaks. ESPIKE~\cite{52} is another fuzzing tool, which is an extension of SPIKE; it is designed to handle encrypted protocols by sending all the SPIKE data through the SSL layer. The limitation of SPIKE is that it is only valid for the protocols that are already compatible with SPIKE.

In general, the main challenge of fuzzing is that it is not directly applicable to encryption because exchanging encrypted messages include randomly generated numbers or symbols~\cite{27}. Considering that mutating encrypted messages always lead to a non-understandable format, the fuzzer needs to have access to the encryption key and the cryptographic algorithm used in the protocol. Thus, in our proposed \textit{EBF} method, the server automatically transmits the information required for decryption (i.e., encryption keys and the cryptographic algorithm used in the protocol) from a log file to the fuzzer. State-of-the-art methods employed only fuzzing or symbolic execution for cryptographic network protocols; none of them ever considered BMC techniques, which outperform typical path-based symbolic execution~\cite{73}. Given the current knowledge in software verification and applied cryptography, this is the first method that combines BMC and fuzzing to detect software vulnerabilities for cryptographic protocols in IoT. We expect by designing this algorithm that we efficiently cover most code paths and avoid the path explosion issue. Moreover, we intend to develop an automated method that finds the most common kinds of vulnerabilities.

As future work, we will implement our entire methodology. In the beginning, we will evaluate ESBMC in the TLS library and validate any bugs ESBMC finds, based on witness checking and later confirmation by developers. Then, we will further develop and evaluate AFL when two endpoints are sharing encrypted messages. Last, we will create the log file to automatically provide AFL with the encryption key ad algorithm used and report the result.

\bibliographystyle{ACM-Reference-Format}
\bibliography{acmart}

\end{document}